\documentclass[a4paper,10pt]{article}
\usepackage[T1]{fontenc}      % codifica dei font in uscita
\usepackage[utf8]{inputenc}   % lettere accentate da tastiera
\usepackage[english]{babel}   % lingua principale del documento
\usepackage{url}              % per scrivere gli indirizzi Internet
\usepackage{bm}
\usepackage{amsmath}
\usepackage{amsfonts}
\usepackage{graphicx}
\usepackage[]{appendix}
\usepackage{listings,lstautogobble}

\usepackage{tabularx}
\usepackage{pdflscape}
\usepackage{longtable}
\usepackage{bigints}
\usepackage{amsmath}
\usepackage{amssymb}
\usepackage{float}
\usepackage[small,bf]{caption2}
\usepackage{bm}
\usepackage{geometry}
\usepackage{etoolbox}
\usepackage{amsthm}
\usepackage{amsmath}
\usepackage{xcolor}

\usepackage{geometry}
\usepackage{graphicx}
\graphicspath{{img/}}

\usepackage{scalerel}
\usepackage{xcolor}

\usepackage{multirow}

\definecolor{blucite}{RGB}{12,127,172}
\usepackage[authoryear]{natbib}
\usepackage[colorlinks, citecolor=blucite]{hyperref}

\DeclareMathOperator*{\argmax}{arg\,max}

\begin{document}
	
	\title{\textbf{Longitudinal Latent Overall Toxicity (LOTox) profiles in osteosarcoma: a new taxonomy based on latent Markov models}\\ \vspace{5mm}}
	
	\author{Marta Spreafico$^{*,1,2,3}$\hspace{1cm} Francesca Ieva$^{1,3,4}$ \hspace{1cm} Marta Fiocco$^{2,5,6}$
	\\ \quad \\
	\small $^1$MOX -- Department of Mathematics, Politecnico di Milano, Milan 20133, Italy\\
	\small $^2$Mathematical Institute, Leiden University, Leiden, The Netherlands\\
	\small $^3$CHRP -- National Center for Healthcare Research and Pharmacoepidemiology, Milan 20126, Italy\\
	\small $^4$CHDS -- Center for Health Data Science, Human Technopole, Milan 20157, Italy\\
	\small $^5$Department of Biomedical Data Sciences, Leiden University Medical Center,  Leiden, The Netherlands\\
	\small $^6$Trial and Data Center, Princess M\'{a}xima Center for Pediatric Oncology, Utrecht, The Netherlands \\ \quad}
	\date{ *\texttt{m.spreafico@math.leidenuniv.nl}  }

	\maketitle
	%\vspace{5mm}
	
	\begin{abstract}
				Due to the presence of multiple types of adverse events with different levels of severity, the analysis of longitudinal toxicity data is a difficult task in cancer studies. In this work, a novel approach based on latent Markov models and compositional data techniques is proposed. The latent status of interest is the Latent Overall Toxicity (LOTox) condition of each patient. The main objectives consist in identifying different latent states of overall toxicity burden and investigating the evolution of individual toxicity risk during cancer treatment.  This methodology is applied to osteosarcoma treatment data to provide novel techniques that may support medical decisions in childhood cancer therapy.
	\end{abstract}
	
		\vspace{0.5cm}
	
\section{Introduction}
\label{s:intro}

Osteosarcoma is a malignant bone tumour mainly affecting children and young adults. Although osteosarcoma is the most common primary malignant bone cancer, it is a rare disease and has an annual incidence of 3-4 patients per million (\citealp{smeland2019}). Multidisciplinary management including neoadjuvant and adjuvant chemotherapy with aggressive surgical resection (\citealp{ritter2010}) or intensified chemotherapy has improved clinical outcomes but over the past 40 years there have been no further improvements in survival (\citealp{anninga2011}).

In cancer trials, the relationship between chemotherapy dose and clinical efficacy outcomes are problematic to analyse due to the presence of negative feedback between exposure to cytotoxic drugs and other aspects, such as latent accumulation of chemotherapy-induced toxicity. Toxicities, developed by patients through chemotherapy, are time-dependent confounders for the effect of chemotherapy on patient’s status (\citealp{lancia2019}). Toxicities affect subsequent exposure by delaying the next cycle or reducing chemotherapy doses (\citealp{souhami1997}), being at the same time risk factors for mortality and predictors for future exposure levels. According to the Common Terminology Criteria for Adverse Events (CTCAE) (\citealp{ctcae3}), a  multimodality grading system for the standardized classification of adverse events (AEs) in cancer therapy, nominal grades of AEs severity range from minor  to life-threatening injuries or death (\citealp{trotti2003}). Since patients may have multiple AEs with different levels of severity, identifying the actual extent of toxic burden  and investigating the evolution of patient's overall toxicity status during treatment represent challenging problems in cancer research. 

Due to the complexity of longitudinal chemotherapy data, no standard method is available for analysing AEs data. Toxicity data are mainly used in cancer studies as summary indexes, such as maximum toxicity over time, maximum grade among events, or weighted sums of individual toxic effects (\citealp{bekelethall2004,rogatko2004,trotti2007,lee2012,mctiernan2012,sivendran2014,thanarajasingam2015,thanarajasingam2016,zhang2016,carbini2018}). Although these methods can summarise data over time, substantial amount of information (e.g., isolated vs repeated events, single vs multiple episodes, longer-lasting lower-grade toxicities, toxic events timing) are discarded. As neglecting the time component may give an inaccurate depiction of toxicity, alternative methods for a longitudinal analysis of AEs have been proposed (\citealp{trotti2007,thanarajasingam2016,thanarajasingam2020,hirakawa2019,spreaficoBMJOpen}). These approaches are not suitable for the nominal CTCAE grades still they provide more insights into treatment-related toxicity, suggesting that longitudinal methods should become routine in future analyses of cancer trials. Models to deal with both longitudinal and categorical aspects of toxicity levels progression are then necessary, still not well developed.

Longitudinal data are often of interest in a wide range of research fields, such as social, economic and behavioural sciences, education or public health. 
In many applications involving longitudinal data, the interest lies in analysing the evolution of a latent characteristic of a group of individuals over time, rather than in studying their observed attributes (\citealp{bartolucci2014}). The phenomenon which affects the distribution of the response variables that are relevant for the problem under consideration may not be directly observable. In a clinical context, this latent characteristic may reflect patients' quality-of-life and could contain valuable information related to patient's health status and disease progression. 

In the statistical literature many models have been proposed for the analysis of longitudinal data; for a concise review see \cite{fitzmaurice2009}. 
For longitudinal categorical data, where the interest is in describing individual changes with respect to a latent status, Latent Markov (LM) models can be used (\citealp{wiggins1973,bartolucci2013book}).
These models study the evolution of an individual characteristic of interest, when it is not directly observable. The idea behind a LM model is that the latent process fully explains the observable behaviour of a subject, assuming that the response variables are conditionally independent given the latent process. The latent process follows a Markov chain with a finite number of states, which represent different conditions of the latent characteristic of interest. LM models can also account for the effect of observable covariates, serial dependence between observations, measurement errors, or unobservable heterogeneity. For a detailed overview on LM models see \cite{bartolucci2013book,bartolucci2014}.

Motivated by the need to improve methods for summarising and quantifying the overall toxicity level and its evolution during treatment, in this work a novel procedure based on LM models for longitudinal toxicity data is proposed. The latent status of interest is the Latent Overall Toxicity (LOTox) condition of a patient, which affects the distribution of the observed categorical toxic grades measured over treatment.
The proposed approach aims at identifying different latent states of overall toxicity burden (\textit{LOTox states}) and investigating how patients move between states during chemotherapy treatment. 

A LM model for longitudinal toxicity data assumes that at each time occasion  for each patient a vector of probabilities of being in the various LOTox states is given. Since the probability elements of each vector are non-negative coordinates whose sum is one, these vectors are naturally confined to a suitably dimensioned simplex, thus being \textit{Compositional Data} (CoDa) or \textit{compositions}. In statistics, CoDa are quantitative descriptions of the parts of some whole, carrying relative information. In this context, \citet{aitchison1986} developed a methodology based on log-ratio transformations of CoDa, which nowadays represent the mainstream approach in the analysis of compositions formed by probabilities or percentages. Among the developed transformations, the \textit{additive log-ratios} consider a specific reference part in contrast with all the others. In this work, this approach is exploited to compare over time a reference “good" overall toxicity condition (i.e., the LOTox state characterized by the lowest toxicity burden) in contrast with all the other LOTox states, characterized by worsening overall toxicity. In this way, the dynamic risk of experiencing “worse" overall toxicity statuses relative to a “good" toxic condition over time is investigated.

Three are the main novelties presented in this work: 
(i) the introduction of a new method based on LM models to summarize and quantify multiple AEs and their evolution during treatment, where both longitudinal and categorical aspects of the observed toxic levels are included in the model; 
(ii) the identification of groups of patients with a common distribution for the observed toxic categories, and thus a similar overall toxicity burden;
(iii) the reconstruction of personalized \textit{longitudinal LOTox profiles}, which represent the probability over time of being in a specific LOTox state or the relative risk with respect to a reference “good" toxic condition, to study the individual overall toxic risk evolution during treatment for each subject. 
The proposed approach is applied to osteosarcoma patients who have completed chemotherapy treatment to provide novel techniques that could support clinicians in planning new protocols and guidelines for childhood cancer therapy. 
Studying the overall toxicity burden in patients who have been able to overcome all the difficulties posed by chemotherapy and reconstructing their longitudinal evolution during treatment may shed some light on a complex aspect of this rare disease.
Provided that longitudinal CTCAE-graded toxicity data are available, the developed procedure is a flexible approach that can be adapted and applied to other cancer studies.

The article is organized as follows. Data from the MRC BO06/EORTC 80931 Randomized Controlled Trial for patients with osteosarcoma (\citealp{lewis2007}) are described in Section \ref{s:data}. Statistical methods are introduced in Section \ref{s:methods}. Results for MRC BO06 data are presented in Section \ref{s:results}. Section \ref{s:discuss} ends with a discussion of strengths and limitations of the proposed approach, identifying some possible developments for future research.

\section{MRC BO06 randomized clinical trial data}
\label{s:data}
In Section \ref{s:data:cohort} the selected cohort of patients is illustrated. Longitudinal chemotherapy data and patient characteristics are presented in Section \ref{s:data:long}.

\subsection{Data illustration} \label{s:data:cohort}
Data from the MRC BO06/EORTC 80931 randomized clinical trial for patients with non-metastatic high-grade osteosarcoma recruited between 1993 and 2002 were analysed (\citealp{lewis2007}). Patients were randomized between conventional (\textit{Reg-C}) and dose-intense (\textit{Reg-DI}) regimens. Both arms had six cycles of the same course of doxorubicin and cisplatin with different time schedule (3-weekly vs 2-weekly, supported by granulocyte colony stimulating factor). Details concerning the trial protocol are provided in Appendix \ref{app:protocol}.

\begin{figure}
	\centering
	\includegraphics[width=12cm]{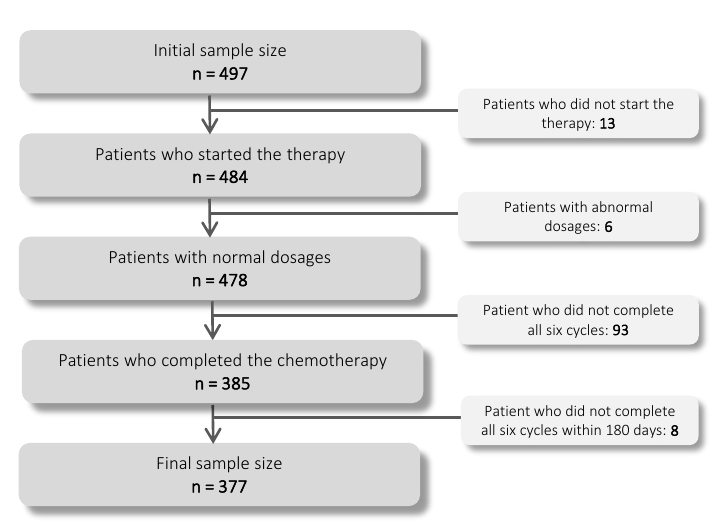}
	\caption{\label{fig:consort} Flowchart of cohort selection.}
\end{figure}

Based on clinical guidance, osteosarcoma patients who had successfully completed the six-cycle treatment were selected for this study.
The initial dataset included 497 eligible patients; 19 patients who did not start chemotherapy (13) or reported an abnormal dosage (i.e., +25\% higher than planned) for a single or both agents (6) were excluded. Patients who did not complete all six cycles of chemotherapy (93) and did not terminate the last cycle within 180 days after randomization (8) were excluded. The final cohort of 377 patients included in the analyses (75.9\% of the initial sample) is shown in the consort diagram in Figure \ref{fig:consort}.

\subsection{Longitudinal chemotherapy data}\label{s:data:long} 
During the trial treatment, case report forms were used to document across cycles all the information required by the  MRC BO06/EORTC 80931 trial protocol for each patient.

Patients baseline characteristics (age, gender, allocated chemotherapy regimen, site and location of the tumour) were registered at randomization.
Among 377 patients, 229 (60.7\%) were males and \textit{Reg-DI} was allocated in 52.3\% of the patients (197). Median age was 15 years (IQR [11; 18]). 
Treatment-related factors (administered dose of chemotherapy, cycles delays,  chemotherapy-induced toxicity, haematological parameters) were collected at each cycle of chemotherapy. For each patient $i$ and cycle $t$, chemotherapy dose was analysed as percentage of achieved chemotherapy dose up to cycle $t$, i.e., the percentage of the cumulative drugs administrated up to cycle $t$ divided by the cumulative drugs planned up to $t$.
Non-haematological chemotherapy-induced toxicity for nausea/vomiting ($naus$), infection ($inf$), oral mucositis ($oral$), cardiac toxicity ($car$), ototoxicity ($oto$) and neurological toxicity ($neur$) were graded according to the Common Terminology Criteria for Adverse Events Version 3 (CTCAE v3.0) (\citealp{ctcae3}), with grades ranging from 0 (none) to 4 (life-threatening) (see Appendix \ref{app:protocol} for further details). Nausea/vomiting, infection and oral mucositis were classified as \textit{generic} toxicities since they represent common adverse events for chemotherapeutic treatments in general. Cardiac toxicity, ototoxicity and neurological toxicity, which could also cause irreversible conditions (see Appendix Table \ref{t:ctcae}), were classified as \textit{drug-specific} toxicities since they are related to the use of cisplatin or doxorubicin (\citealp{DOXcardiotox,CDDPtox}).

Considering the CTCAE-grades registered over cycles for each non-haematological adverse event, \textit{generic} toxicities were more frequent than  \textit{drug-specific} ones, as expected.
Nausea/vomiting was reported at least once over cycles in 97.3\% of patients (367/377), with a percentage that decreased over cycles from 84.9\% in cycle 1 to 52.5\% in cycle 6. The percentages of patients who reported oral mucositis or infections were more stable over cycles: 30.5\%--43.3\% for mucositis, with 78\% (294/377) reporting mucositis at least once, and 23.8\%--31.3\% for infection, with 69\% (260/377) reporting an infection at least once. Ototoxicity was reported at least once in 21.5\% (81/377), cardiac toxicity in 14.1\% (53/377) and neurological toxicity in 11.7\% (44/377). At each cycle, CTCAE-grade 4 for \textit{generic} toxicities and CTCAE-grades $\ge2$ for \textit{drug-specific} toxicities were reported in less than 5\% of patients. Low-frequency classes were merged and toxic categories were represented according to the degree of severity or as present or not, depending on the type of toxicity as follows:
\begin{itemize}
	\item the severity of the toxic event for \textit{generic} toxicities: \textit{none} (CTCAE-grade 0), \textit{mild} (CTCAE-grade 1), \textit{moderate} (CTCAE-grade 2), and \textit{severe} (CTCAE-grades 3 or 4);
	\item the absence or the presence of toxic event for \textit{drug-specific} toxicities: \textit{no} (CTACE-grade 0) and \textit{yes} (CTACE-grades $\ge1$).
\end{itemize} 
These categories identified for each toxicity constitute the item-response elements selected to model the latent process representing the “true” overall toxic status. Table \ref{t:toxcat} shows the observed frequencies (and percentages) of the selected categories for each toxicity over cycles for the final cohort. The observed responses for each patient are then given by the longitudinal toxic categories measured along the cycles, which are  used to evaluate the LOTox condition during treatment.

\begin{table}[t]
	\caption{\label{t:toxcat} Frequencies of toxic categories over the six cycles. For nausea, infection and mucositis $(j=1,2,3)$, the set of toxic categories indicating the severity of the toxic event is defined as $\mathcal{C}_j=\{none;\, mild;\, moderate;\, severe\}$. For cardiotoxicity, otoxocity and neurological toxicity  $(j=4,5,6)$, the set of toxic categories indicating the presence or the absence of the toxic event is defined as $\mathcal{C}_j=\{no;\, yes\}$. }
	\small
	\centering
	\begin{tabular}{lrrrrrr}
		\hline
		Toxicity & \multicolumn{1}{c}{Cycle 1} & 
		\multicolumn{1}{c}{Cycle 2} & \multicolumn{1}{c}{Cycle 3} & 
		\multicolumn{1}{c}{Cycle 4} & 
		\multicolumn{1}{c}{Cycle 5} & 
		\multicolumn{1}{c}{Cycle 6}\\ \hline
		
		\multicolumn{2}{l}{\textbf{Nausea}} &&&&&\\
		~~\textit{none} & 57 (15.1\%) & 88 (23.3\%) & 115 (30.5\%) & 126 (33.4\%) & 146 (38.7\%) & 179 (47.5\%) \\ 
		~~\textit{mild} & 74 (19.6\%) & 87 (23.1\%) & 76 (20.2\%) & 72 (19.1\%) & 86 (22.8\%) & 74 (19.6\%) \\ 
		~~\textit{moderate} & 117 (31.1\%) & 117 (31.1\%) & 114 (30.2\%) & 113 (30.0\%) & 96 (25.5\%) & 87 (23.1\%) \\ 
		~~\textit{severe} & 129 (34.2\%) & 85 (22.5\%) & 72 (19.1\%) & 66 (17.5\%) & 49 (13.0\%) & 37 ~(9.8\%) \\ \hline
		
		\multicolumn{2}{l}{\textbf{Infection}} &&&&&\\
		~~\textit{none} & 259 (68.7\%) & 287 (76.1\%) & 268 (71.1\%) & 265 (70.3\%) & 268 (71.1\%) & 286 (75.9\%) \\ 
		~~\textit{mild} & 30 \, (7.9\%) & 24 \, (6.4\%) & 26 \, (6.9\%) & 31 \, (8.2\%) & 23 \, (6.1\%) & 16 \, (4.3\%) \\ 
		~~\textit{moderate} & 64 (17.0\%) & 45 (11.9\%) & 61 (16.2\%) & 54 (14.3\%) & 52 (13.8\%) & 45 (11.9\%) \\ 
		~~\textit{severe} & 24 \, (6.4\%) & 21 \, (5.6\%) & 22 \, (5.8\%) & 27 \, (7.2\%) & 34 \, (9.0\%) & 30 \, (8.0\%) \\ \hline
		
		\multicolumn{2}{l}{\textbf{Mucositis}} &&&&&\\
		~~\textit{none} & 265 (70.3\%) & 228 (60.5\%) & 234 (62.1\%) & 237 (62.9\%) & 214 (56.8\%) & 262 (69.5\%) \\ 
		~~\textit{mild} & 54 (14.3\%) & 46 (12.2\%) & 59 (15.6\%) & 52 (13.8\%) & 62 (16.4\%) & 44 (11.7\%) \\ 
		~~\textit{moderate} & 44 (11.7\%) & 54 (14.3\%) & 43 (11.4\%) & 55 (14.6\%) & 63 (16.7\%) & 50 (13.2\%) \\ 
		~~\textit{severe} & 14 \, (3.7\%) & 49 (13.0\%) & 41 (10.9\%) & 33 \, (8.7\%) & 38 (10.1\%) & 21 \, (5.6\%) \\ \hline
		
		\multicolumn{2}{l}{\textbf{Cardiotoxicity}} &&&&&\\
		~~\textit{no} & 374 (99.2\%) & 361 (95.8\%) & 362 (96.0\%) & 359 (95.2\%) & 357 (94.7\%) & 355 (94.2\%) \\ 
		~~\textit{yes} & 3 \, (0.8\%) & 16 \, (4.2\%) & 15 \, (4.0\%) & 18 \, (4.8\%) & 20 \, (5.3\%) & 22 \, (5.8\%) \\ \hline
		
		\multicolumn{2}{l}{\textbf{Ototoxicity}} &&&&&\\
		~~\textit{no} & 357 (94.7\%) & 361 (95.8\%) & 350 (92.8\%) & 342 (90.7\%) & 346 (91.8\%) & 326 (86.5\%) \\ 
		~~\textit{yes} & 20 \, (5.3\%) & 16 \, (4.2\%) & 27 \, (7.2\%) & 35 \, (9.3\%) & 31 \, (8.2\%) & 51 (13.5\%) \\ \hline
		
		\multicolumn{2}{l}{\textbf{Neurological toxicity}} &&&&&\\
		~~\textit{no} & 371 (98.4\%) & 367 (97.3\%) & 362 (96.0\%) & 367 (97.3\%) & 356 (94.4\%) & 363 (96.3\%) \\ 
		~~\textit{yes} & 6 \, (1.6\%) & 10 \, (2.7\%) & 15 \, (4.0\%) & 10 \, (2.7\%) & 21 \, (5.6\%) & 14 \, (3.7\%) \\ 
		
		\hline
	\end{tabular}
\end{table}

\section{Statistical Methods}
\label{s:methods}
In the following sections, the novel Latent Markov (LM) approach for modelling the Latent Overall Toxicity (LOTox) condition of each patient starting from the observed longitudinal toxic categories measured during chemotherapy treatment is introduced.
In Section \ref{s:methods:lmmot} motivations for the proposed approach for treating the longitudinal toxicity data are discussed.
Mathematical details are provided in Section \ref{s:methods:model}. Model selection procedure and longitudinal profiles are presented in Sections \ref{s:methods:sel} and \ref{s:methods:prof}, respectively.

\subsection{Motivations for latent Markov models for longitudinal toxicity data}\label{s:methods:lmmot} 
LM models are statistical methods employed for the analysis of longitudinal (categorical) data specifically designed to study the evolution of an individual characteristic of interest, when it is not directly observable (\citealp{wiggins1973,bartolucci2013book}).
A LM approach for longitudinal toxicity data assumes the existence of a latent process representing the “true" LOTox status, which affects the distribution of the response variables, in our case the observed toxicities. Two main motivations justify the use of LM models to quantify the toxic risk in cancer studies: (i) account for \textit{measurement errors} in the observed toxicity variables, and (ii) identify different \textit{LOTox sub-populations} (i.e., the latent states) in the global population (i.e., the patients' cohort) and their changes over time.  

Since therapy protocol is adapted at each cycle depending on patient's reaction to treatment, it is reasonable to assume that the latent variables follow a first-order Markov chain, so that the “true" level of overall toxicity at a given cycle is influenced only by the previous level. Non-hematological toxicities  do not depend directly on each other as they relate to different systems and functions of the human body (i.e., nausea/vomiting is part of the stomach-gastrointestinal system, infections of the immune system, oral mucositis of the mouth-gastrointestinal system, cardiotoxicity of the cardiovascular system, ototoxicity of the auditory-sensory system and neurotoxicity of the nervous system). Therefore, the response toxicity variables can be assumed conditionally independent, as each observed response is expected to depend only on the corresponding “true" LOTox level. 

In this context, a LM model may be seen as an extension of the latent class model (\citealp{collinslanza2010}), where patients are allowed to move between latent states during the observation period.
LM models for longitudinal toxicity data are characterized by several parameters: the initial probability of each LOTox state, the transition probabilities among different states over chemotherapy cycles, and the conditional response probabilities given the latent variable. Individual covariates (if available) can be included in the latent model and may affect the initial and transition probabilities of the Markov chain (\citealp{bartolucci2009,bartolucci2013book}), as explained in Section \ref{s:methods:model}.

A LM approach is appropriate to both identify the actual overall toxicity burden and investigate its evolution during treatment for each patient.
On one hand, patients that at a specific time result in the same sub-population are characterized by a common distribution for the observed toxic categories, and by a similar overall toxicity burden. On the other hand, individual dynamic changes among latent states allow to evaluate the LOTox evolution during treatment for each subject.

\subsection{Latent Markov model with covariates}\label{s:methods:model}
Let $\mathcal{J}$ be the set of $J=|\mathcal{J}|$ categorical response variables measured at each time $t = 1, \dots, T$. 
Denote by $Y_{ij}^{(t)}$ the response variable $j \in  \{1, \dots, J\}$ for subject $i \in \{1,...,n\}$ at time $t$, with set of categories $\mathcal{C}_j$ coded from 0 to $c_j-1$.
Let $\bm{Y}_i^{(t)} = \left(Y_{i1}^{(t)},...,Y_{iJ}^{(t)}\right)$ denote the observed multivariate response vector at time $t$ for patient $i$ and $\bm{\widetilde Y}_i = \left(\bm{Y}_i^{(1)},\dots, \bm{Y}_i^{(T)}\right)$ be the corresponding complete response vector.
Denote by $\bm{\widetilde X}_i = \left(\bm{X}_i^{(1)},\dots, \bm{X}_i^{(T)}\right)$ the complete vector of individual covariates, where elements $\bm{X}_i^{(t)}=\left(\bm{S}_i,\bm{Z}_i^{(t)}\right)$ are the vectors of time-fixed $\bm{S}_i$ and time-varying $\bm{Z}_i^{(t)}$ covariates for subject $i$ at occasion $t$.
The general LM model assumes the existence of a latent process $\bm{U}_i = \left(U_i^{(1)},\dots, U_i^{(T)}\right)$ which affects the distribution of the response variables $\bm{\widetilde Y}_i$. The latent process follows a first-order Markov chain with state space $\{1, \dots, k\}$, where $k$ is the total number of \textit{latent states}. LM models usually assume that the response vectors $\bm{Y}_i^{(1)},\dots, \bm{Y}_i^{(T)}$ are conditionally independent given the latent process $\bm{U}_i$ (\textit{local independence of the response vectors}) and that the elements $Y_{ij}^{(t)}$ are conditionally independent given $U_i^{(t)}$ (\textit{conditional independence of elements}). The motivation of these assumptions is that the latent process fully explains the observable behaviour of a subject, as explained in Section \ref{s:methods:lmmot}.

LM models are made by two components: the \textit{measurement model} concerns the conditional distribution of the response variables given the latent process, and the \textit{latent model} is related to the distribution of the latent process (i.e., initial and transition probabilities).
The latent process represents an individual characteristic of interest that is not directly observable that may evolve over time, also depending on observable covariates. The main research interest hence lies in modelling the latent process and the effect of covariates on its dynamic.
LM models where both the initial and the transition probabilities of the latent process may depend on covariates is considered.  Three different sets of probabilities (i.e., parameters) can be defined.

\begin{itemize}
	\item \textit{Conditional response probability} (or item-response probability) $\phi^{(t)}_{jy|u}$ is the probability of observing a response $y$ for variable $j$ at time $t$, given the latent status $u\in \{1,...,k\}$:
	\begin{equation*}
		\mathrm{P}\left(Y_{ij}^{(t)}=y\big|U_i^{(t)}=u\right)=\phi^{(t)}_{jy|u} \qquad j=1,\dots,J \quad y=0,...,c_j-1. 
	\end{equation*}
	To ensure that the interpretation of the latent states remains constant over time, conditional response probabilities are assumed time-homogeneous, i.e., $\phi^{(t)}_{jy|u}=\phi_{jy|u}$ $\forall  t=1,\dots,T$. Given the estimated $\hat\phi_{jy|u}$, the latent states can be characterized in terms of observed response categories.
	\item \textit{Initial latent states prevalence}  $\delta_{u|\bm{x}_i^{(1)}} $ is the probability of membership in latent state $u  \in \{1, \dots, k\}$ at time $t=1$, given the vector of covariates $\bm{x}_i^{(1)}$ for individual $i$:
	\begin{equation*}
		\mathrm{P}\left(U_i^{(1)}=u | \bm{X}_i^{(1)}= \bm{x}_i^{(1)} \right) = \delta_{u|\bm{x}_i^{(1)}}.
	\end{equation*}
	The estimated  $\hat \delta_{u|\bm{x}_i^{(1)}}$ may be interpreted as quantities proportional to the size of each latent state at the first time-occasion, given the covariates. A natural way to allow the initial probabilities of the LM chain to depend on individual covariates is a multinomial logit parametrization:
	\begin{equation}\label{eq:betas}
		\log \frac{\mathrm{P}\left(U_i^{(1)}=u \mid \boldsymbol{X}_i^{(1)}=\boldsymbol{x}_i^{(1)}\right)}{\mathrm{P}\left(U^{(1)}=1 \mid \boldsymbol{X}_i^{(1)}=\boldsymbol{x}_i^{(1)}\right)}=\log \frac{\delta_{u \mid \boldsymbol{x}_i^{(1)}}}{\delta_{1 \mid \bm{x}_i^{(1)}}}=\beta_{0 u}+\boldsymbol{x}_i^{{(1)}\top} \boldsymbol{\beta}_{1 u}
	\end{equation}
	where $u=2,...,k$ and $\boldsymbol{\beta}_{u} = \left(\beta_{0u}, \boldsymbol{\beta}_{1 u}^{\top} \right)^{\top}$ are the parameters vectors to be estimated.
	
	\item \textit{Transition probability} $\tau_{u|\bar{u}\bm{x}_i^{(t)}}^{(t)}$ is the probability of a transition to latent state $u$ at time $t$, conditional on membership in latent state $\bar{u}$ at time $t-1$, given the individual vector of covariates $\bm{x}_i^{(t)}$ (if available):
	\begin{equation*}
		\mathrm{P}\left(U_i^{(t)}=u\mid U_i^{(t-1)}=\bar u, \bm{X}_i^{(t)}= \bm{x}_i^{(t)} \right) = \tau_{u \mid \bar{u}\bm{x}_i^{(t)}}^{(t)}
	\end{equation*}
	where $t=2,\dots,T$ and $u,\bar{u} =1, \dots, k$. The estimated $\hat \tau_{u \mid \bar{u}\bm{x}_i^{(t)}}^{(t)}$ reflect changes or persistence in the various states over time, given the individual covariates whose effects can be modelled through a multinomial logit parametrization:
	\begin{equation}\label{eq:gammas}
		\log \frac{\mathrm{P}\left(U_i^{(t)}=u \mid U_i^{(t-1)}=\bar{u}, \boldsymbol{X}_i^{(t)}=\boldsymbol{x}_i^{(t)}\right)}{\mathrm{P}\left(U_i^{(t)}=\bar{u} \mid U_i^{(t-1)}=\bar{u}, \boldsymbol{X}_i^{(t)}=\boldsymbol{x}_i^{(t)}\right)}=\log \frac{\tau_{u \mid \bar{u} \bm{x}_i^{(t)}}^{(t)}}{\tau_{\bar{u} \mid \bar{u} \bm{x}_i^{(t)}}^{(t)}}=\gamma_{0 \bar{u} u}+\boldsymbol{x}_i^{(t)\top} \boldsymbol{\gamma}_{1 \bar{u} u}
	\end{equation}
	for  $t = 2,...,T$ and $\bar{u},u = 1, ...,k$ with $\bar{u}\neq u$.  $\boldsymbol{\gamma}_{\bar u u} = \left(\gamma_{0\bar u u}, \boldsymbol{\gamma}_{1 \bar u u}^{\top} \right)^{\top}$ are the parameters vectors to be estimated.
\end{itemize}

Under the assumptions of \textit{local}  and \textit{conditional independence}, the \textit{manifest distribution} of the response variables (i.e., the conditional distribution of $\bm{\widetilde Y}_i$ given $\bm{\widetilde X}_i$) is given by:
\begin{equation}\label{eq:manifest}
	\begin{split}
		\mathrm{P}(\bm{\widetilde y}_i \mid \bm{\widetilde x}_i) &= \mathrm{P}\left(\bm{\widetilde Y}_i=\bm{\widetilde y}_i \mid \bm{\widetilde X}_i=\bm{\widetilde x}_i\right) =\\
		&= \sum_{\bm{u}}^{} \mathrm{P}\left(\bm{\widetilde Y}_i=\bm{\widetilde y}_i \mid \bm{\widetilde X}_i=\bm{\widetilde x}_i, \bm{U}_i=\bm{u}\right) \times \mathrm{P}\left(\bm{U}_i=\bm{u} \mid \bm{\widetilde X}_i=\bm{\widetilde x}_i\right) =\\
		&= \sum_{\bm{u}}^{} \mathrm{P}\left(\bm{U}_i=\bm{u} \mid \bm{\widetilde X}_i=\bm{\widetilde x}_i\right) \times \mathrm{P}\left(\bm{\widetilde Y}_i=\bm{\widetilde y}_i \mid \bm{U}_i=\bm{u}\right)  =\\
		&= \sum_{\bm{u}}^{} \delta_{u^{(1)}\mid \bm{x}_i^{(1)}} \prod_{t=2}^{T} \tau^{(t)}_{u^{(t)} \mid u^{(t-1)} \bm{x}_i^{(t)}}\times \prod_{t=1}^{T} \prod_{j=1}^{J} \phi_{jy_{ij}^{(t)}\mid u^{(t)}}
	\end{split}
\end{equation}
where $\bm{u} = (u^{(1)},\dots, u^{(T)})$. The vector $\bm{\widetilde y}_i=\left(\bm{y}_i^{(1)},\dots, \bm{y}_i^{(T)}\right)$ is a realization of $\bm{\widetilde Y}_i$, where $\bm{y}_i^{(t)}$ is an observation of $\bm{Y}_i^{(t)}$ with elements $y_{ij}^{(t)}$. 
The vector $\bm{\widetilde x}_i=\left(\bm{x}_i^{(1)},\dots, \bm{x}_i^{(T)}\right)$ is a realization of $\bm{\widetilde X}_i$, where $\bm{x}_i^{(t)}=\left(\bm{s}_i,\bm{z}_i^{(t)}\right)$ is an observation of $\bm{X}_i^{(t)}=\left(\bm{S}_i,\bm{Z}_i^{(t)}\right)$.

Parameter estimation is performed maximizing the log-likelihood
\begin{equation}\label{eq:loglik} 
	\ell(\bm{\theta})=\sum_{i=1}^{n} \log P(\bm{\widetilde y}_i \mid \bm{\widetilde x}_i)
\end{equation}
where $\bm{\theta}$ denotes the complete parameter vector.  The log-likelihood \eqref{eq:loglik} can be maximized by Expectation-Maximization (EM) algorithm (\citealp{Baum1970,Dempster1997}), as explained in the following section. Alternative algorithms are also available but they are more difficult to implement and less stable (\citealp{bartolucci2013book}).

\subsubsection{Expectation-Maximization algorithm}
The EM algorithm is based on the complete data log-likelihood (\citealp{bartolucci2013book,bartolucci2014}). For multivariate categorical data, a computationally more convenient expression equivalent to \eqref{eq:loglik} is used which is based on the joint frequency of the covariate and response configurations:
\begin{equation}\label{eq:loglik2}
	\begin{aligned}
		\ell^*(\boldsymbol{\theta})=
		\sum_{j=1}^J \sum_{t=1}^T \sum_{u=1}^k \sum_{\bm{x}} \sum_{y=0}^{c_j-1} a_{j u \bm{x} y}^{(t)} \log \phi_{j y \mid u}^{(t)}+ & \sum_{u=1}^k \sum_{\bm{x}} b_{u \bm{x}}^{(1)} \log \delta_{u \mid \boldsymbol{x}}+\\
		& \sum_{t=2}^T \sum_{\bar{u}=1}^k \sum_{u=1}^k \sum_{\boldsymbol{x}} b_{\bar{u} u \boldsymbol{x}}^{(t)} \log \pi_{u \mid \bar{u} \boldsymbol{x}}^{(t)}
	\end{aligned}
\end{equation}
where, with reference to occasion $t$ and covariate configuration $\bm{x}$, $a_{j u \bm{x} y}^{(t)}$ is the number of subjects that are in the latent state $u$ and provide response $y$,  $ b_{u \bm{x}}^{(1)}$ is the frequency of latent state $u$, and $b_{\bar{u} u \boldsymbol{x}}^{(t)}$ is the number of transitions from state $\bar{u}$ to state $u$.

The EM algorithm alternates two steps until convergence, which is checked on the basis of the relative log-likelihood difference with respect to a proper tolerance level. The \textit{E-step} consists  in computing the conditional expected value of each frequency involved in the complete data log-likelihood \eqref{eq:loglik2}, given the observed data and the current values of parameters.  The \textit{M-step} updates the estimate of parameter vector $\bm{\theta}$ by separately maximizing the three components of \eqref{eq:loglik2}, with each frequency substituted by the corresponding expected value. A detailed description can be found in \cite{bartolucci2013book}.

As typically happens for latent variable models, the function  $\ell^*(\boldsymbol{\theta})$ may be multimodal and the algorithm may converge to a local mode of the likelihood which could not correspond to the global maximum of  \eqref{eq:loglik2}. In particular, the number of local maxima usually increases as the number of latent states increases and the sample size decreases. To overcome this issue, a multi-start strategy which combines a deterministic rule with a random starting rule can be adopted. For an extensive discussion see \cite{bartolucci2013book,bartolucci2014}.

\subsection{Model selection}\label{s:methods:sel}
The choice of the final LM model for the application consists of two steps: (i) identification of the number of latent states $k$, and (ii) selection of the covariates to be included in the final model.
When the number of latent states $k$ can not be a priori defined based on clinical indications, it can be selected according different measures. Akaike information criterion (AIC) by \cite{akaike1973} or the Bayesian information criterion (BIC) by \cite{schwarz1978}, defined as
\begin{equation*}
	\text{AIC} = - 2 \hat{\ell} + 2g\qquad \text{ and } \qquad
	\text{BIC} = - 2 \hat{\ell} + \log(n)g,
\end{equation*}
where $\hat{\ell}$ is the maximum of the log-likelihood of the model of interest and $g$ denotes the number of free parameters, are used. In particular, the smaller the values of the above criteria, the better the  model represents the optimum compromise between goodness-of-fit and complexity. If the two criteria lead to selecting a different number of states, BIC is usually preferred (\citealp{bacci2014,bartolucci2013book,LMest}).

Basic LM models (i.e., LM models with time-heterogeneous transitions and no covariates - named M1) were fitted increasing the value of $k$ from 1 to 10, and the number of latent states $k$ was selected according to the minimum BIC.
Once $k$ was determined, a forward strategy was adopted to identify the covariates to be included in the final model. In particular, the smallest basic LM model with $k$ latent states and time-homogeneous transitions (i.e., the LM model restricted to the case in which initial and transition probabilities are parametrized by multinomial logit without covariates - named M2) was initially fitted and then the effect of each covariate on initial and/or transition probabilities (models M3-M12) was added. Only the covariates whose effect reduces the value of the BIC index were included in the final LM model.

\subsection{Longitudinal profiles: latent probability and relative risk}\label{s:methods:prof}
In LM models literature, once the model has been estimated, a decoding procedure is usually implemented to obtain a path prediction for each subject, i.e., finding the most likely sequence of latent states on the basis patient-specific observed data (\citealp{bartolucci2013book,bartolucci2014}). 
However, this sequence represents a summary of how the entire latent process evolves over time, as it only provides information about the most-likely condition without giving details about other states (see Appendix \ref{app:decoding}). To obtain more insights into the entire latent process and its evolution, longitudinal information related to each latent state can be reconstructed for each subject.

For each patient-specific observed data $(\bm{\widetilde x}_i,\bm{\widetilde y}_i)$, the Expectation-Maximization algorithm provides the \textit{posterior} probabilities of variables $U_i^{(t)}$
\begin{equation}\label{eq:posterior}
	p_{iu}^{(t)} = \mathrm{P}\left(U_i^{(t)}=u \big| \bm{\widetilde Y}_i=\bm{\widetilde y}_i, \bm{\widetilde X}_i=\bm{\widetilde x}_i\right) \qquad t=1,\dots,T \quad u \in \{1,...,k\},
\end{equation}
which can be estimated using recursions and involving the \textit{manifest }distribution in Equation (\ref{eq:manifest}). 
For each latent state $u\in \{1,\dots,k\}$, probabilities in (\ref{eq:posterior}) can be used to reconstruct the \textit{longitudinal latent probability profile} of the $i$-th subject, as follows:
\begin{equation}\label{eq:llpp}
	\bm{p}_{iu} = \left\{p_{iu}^{(t)} =  \mathrm{P}\left(U_i^{(t)}=u \big| \bm{\widetilde Y}_i=\bm{\widetilde y}_i, \bm{\widetilde X}_i=\bm{\widetilde x}_i\right), \quad t=1,\dots,T \right\}.
\end{equation}
Each profile $\bm{p}_{iu}$ represents the probability over time $t$ of being in latent state $u$ for individual $i$, given the observed complete response $\bm{\widetilde y}_i$ and covariates $\bm{\widetilde x}_i$ (if available). 
Applying this procedure,  $k$ \textit{longitudinal latent probability profiles} (one for each latent state) are obtained for each subject $i$, which can be expressed as a $k \times T$ matrix 
\begin{equation*}
	\bm{P}_i =  \begin{bmatrix}
		\bm{p}_{i1}\\
		\dots\\
		\dots\\
		\bm{p}_{ik}
	\end{bmatrix} = 
	\begin{bmatrix}
		p_{i1}^{(1)} & p_{i1}^{(2)} & \dots & p_{i1}^{(T)}\\
		\dots & & &  \dots\\
		\dots & & &  \dots\\
		p_{ik}^{(1)} & p_{ik}^{(2)} & \dots & p_{ik}^{(T)}
	\end{bmatrix}
	= \begin{bmatrix} \bm{p}_i^{(1)} & \bm{p}_i^{(2)} & \dots & \bm{p}_i^{(T)} \end{bmatrix}
\end{equation*}
with longitudinal latent probability profiles $\bm{p}_{iu}$ as row-components. Columns of $\bm{P}_i$ represent the vectors $\bm{p}_i^{(t)}$ of posterior probabilities over time $t=1,\dots,T$ and can be seen as Compositional Data (CoDa) vectors belonging to the $k$-part Aitchison-Simplex $\mathcal{S}^k$ (\citealp{aitchison1986}), i.e.,
\begin{equation}\label{eq:simplex}
	\bm{p}_i^{(t)} \in \mathcal{S}^k = \left\{ \bm{p} = \left[p_1,...,p_k\right] \in \mathbb{R}^k \Big{|} p_u >0, u=1,\dots,k; \sum_{u=1}^{k}p_u=1\right\}.
\end{equation}
Due to the sum constraint in Equation \eqref{eq:simplex}, elements ${p}_{iu}^{(t)}$ of the composition $\bm{p}_i^{(t)}$ are mutually dependent features which only carry relative information. 
In this context, \citet{aitchison1986} introduced a methodology based on log-ratio transformations of CoDa, which are required to remove constraints and eventually to map the composition to a real space, allowing standard statistical techniques to be applied to the transformed data.
In most practical settings, the choice of transformation will depend on the preferred interpretation.

In the current framework, rather than considering the absolute individual elements ${p}_{iu}^{(t)}$, it is interesting to study the relative risk over time of being in a reference latent state $u = R$ compared to all the other latent states. Among the transformations introduced by \citet{aitchison1986}, this can be done
considering the \textit{additive log-ratios} of each CoDa vector $\bm{p}^{(t)}_i$, as follows: 
\begin{equation}\label{eq:alr}
	\begin{split}
		\mathrm{alr}{\left(\bm{p}^{(t)}_i\right)} &=  \left[\log{\frac{p_{i1}^{(t)}}{p_{iR}^{(t)}}} \, \dots \, \log{\frac{p_{iR-1}^{(t)}}{p_{iR}^{(t)}}} \, \log{\frac{p_{iR+1}^{(t)}}{p_{iR}^{(t)}}} \, \dots \, \log{\frac{p_{ik}^{(t)}}{p_{iR}^{(t)}}}\right]^T\\
		&= \left[r_{i1}^{(t)} \, \dots \, r_{iR-1}^{(t)}  \, r_{iR+1}^{(t)} \, \dots \, r_{ik}^{(t)} \right]^T\\
		&= \bm{r}_i^{(t)} \in \mathbb{R}^{k-1}
	\end{split}
\end{equation} 
where $R$ is the reference latent state which can be chosen arbitrary among $\{1,\dots,k\}$. Note that this transformation maps each bounded sample into a real space $\left(\text{alr: } \mathcal{S}^k \rightarrow \mathbb{R}^{k-1}\right)$ and if one of the $p_{iu}^{(t)}$ elements is exactly zero, a zero-handling procedure is needed before applying the transformation. In that case, an easily applicable possibility would be to replace each zero with a small appropriate value, modifying the non-zero values of the relative composition in a multiplicative way in order to satisfy the sum constraint requirement. For further details see \cite{martinfernandez2011}.
Applying this procedure to each compositions, $k-1$ \textit{longitudinal relative risk profiles} (one for each non-reference state) are obtained for each subject $i$, given as a $(k-1) \times T$ matrix
\begin{equation*}
	\bm{R}_i = \begin{bmatrix} \bm{r}_i^{(1)} & \bm{r}_i^{(2)} & \dots & \bm{r}_i^{(T)} \end{bmatrix} =
	\begin{bmatrix}
		\bm{r}_{i1}\\
		\dots\\
		\bm{r}_{iR-1}\\
		\bm{r}_{iR+1}\\
		\dots\\
		\bm{r}_{ik}
	\end{bmatrix}
\end{equation*}
where column-element $\bm{r}_i^{(t)}$ are given by Equation \eqref{eq:alr} and row-element $\bm{r}_{iu}$ with $u\neq R$ are the \textit{longitudinal relative risk profile} of state $u$ for subject $i$
\begin{equation}\label{eq:lrr}
	\bm{r}_{iu} = \left\{r_{iu}^{(t)} = \log{\frac{p_{iu}^{(t)}}{p_{iR}^{(t)}}}, \quad u\neq R, \, t=1,\dots,T \right\}.
\end{equation}
Each profile $\bm{r}_{iu}$ represents the relative risk (in logarithmic scale) over time $t$ of being in latent state $u \neq R$ with respect to the reference state $R$ for individual $i$. 
Since this procedure is a transformation-based analysis, transformed elements $r_{iu}^{(t)}$ must then be interpreted with respect to the chosen reference. A positive (negative) value $r_{iu}^{(t)}$ at time $t$ means that the risk for subject $i$ of being in latent state $u\neq R$ is $\exp\left\{r_{iu}^{(t)}\right\}$ times higher (lower) than being in reference state $R$.

For the application discussed in this work, the  \textit{LOTox states} summarize different levels of overall toxicity burden, representing a proxy for patient's quality of life. Therefore, for each patient $i$, longitudinal latent probability profile (\ref{eq:llpp}) represents the probability over time of being in the LOTox state $u$ (i.e., the probability over time of developing an overall toxic burden quantified by state $u$) given patient's history: observed toxicity categories $\bm{\widetilde y}_i$ and personal characteristics $\bm{\widetilde x}_i$ over treatment. Once the LOTox states have been identified, it is reasonable to analyse and interpret the different results in relation to the state characterized by the lowest overall toxicity burden (i.e., “good" toxic condition), which is chosen as the reference $R$. In this way, the longitudinal relative risk profile (\ref{eq:lrr}) represents the risk of being in LOTox condition $u \neq R$ compared to the lowest toxic status.

By reconstructing the \textit{longitudinal LOTox profiles}, it is possible to (i) describe patient’s response to therapy over cycles, (ii) quantify the overall toxicity burden evolution over treatment cycles given patient's history and (iii) investigate the individual dynamic changes among latent states, detecting differences in health status and quality of life among patients.

\section{Data application}
\label{s:results}
In this section, the results obtained from the application of the proposed LM model to the MRC BO06/EORTC 80931 randomized clinical trial are reported. Statistical analyses were performed in the \texttt{R}-software environment (\citealp{R}), using \texttt{LMest} package by \cite{LMest}. \texttt{R} code for the current study is available at \url{https://github.com/mspreafico/BO06-LOTox}.

\subsection{Latent Markov model for longitudinal toxicity data}\label{s:results:LM}
For each cycle $t=1,\dots,6$, let $\mathcal{J}=\{naus, \,inf, \,oral, \,car, \,oto, \,neur\}$ be the set of non-haematological toxicities, representing response variables $Y_{ij}^{(t)}$. The relative sets of response categories identified in Section \ref{s:data:long} were coded from 0 to $c_j-1$, as follows:
\begin{align*}
	\mathcal{C}_j &= \{0:none, \,1:mild, \,2:moderate, \,3:severe\} \quad \text{for \textit{generic} toxicities } (j=1,2,3),\\
	\mathcal{C}_j &= \{0:no, \,1:yes\} \quad \text{for \textit{drug-specific} toxicities } (j=4,5,6).
\end{align*}

The procedure described in Section \ref{s:methods:sel} was applied to first identify the number of latent states $k$ and then select the covariates to be included in the final model. Age, gender and allocated regimen at randomization were considered as time-fixed covariates, while percentage of achieved chemotherapy dose up to cycle $t$, white blood cell, neutrophils and platelets counts measured at each cycle were considered  as time-varying ones. 
In order to achieve the global maximum of the log-likelihood \eqref{eq:loglik2}, the multi-start strategy \texttt{R} function \texttt{lmestSearch()}  of  \texttt{LMest} package (\citealp{LMest}), which combines deterministic and random initializations, was exploited for each model.
Specifically, tolerance levels of \texttt{tol1 = 10\textsuperscript{-8} } and \texttt{tol2 = 10\textsuperscript{-10} }  -- for checking convergence of the EM algorithm in the random initializations and in the last deterministic initialization, respectively -- and  \texttt{nrep = 20} repetitions of each random initialization were used.

\begin{figure}[!h]
	\centering
	\includegraphics[width=1\textwidth]{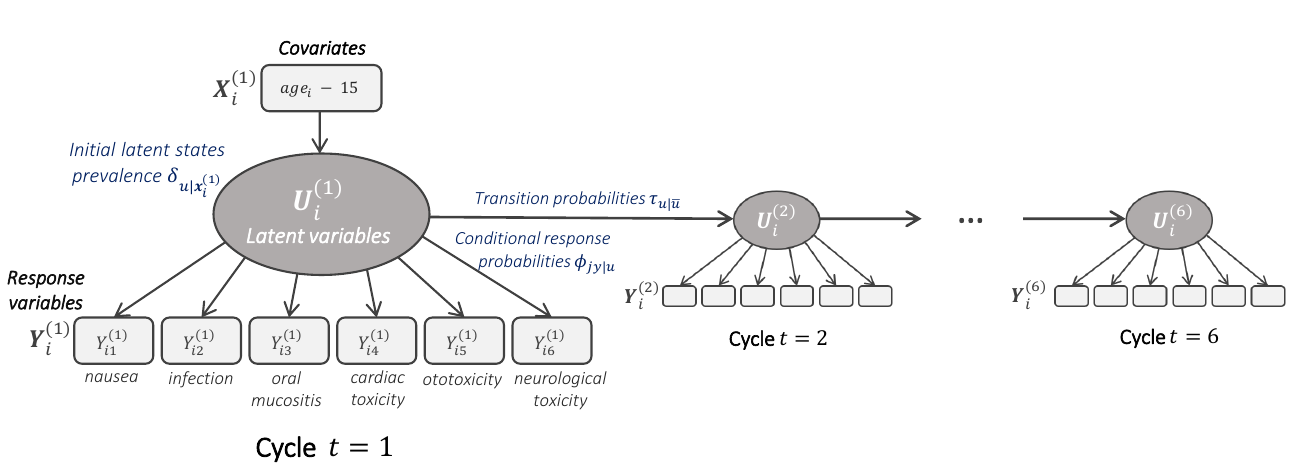}
	\caption{\label{fig:M5}  Path diagram for a given subject $i$ under the latent Markov model M5 with non-haematological toxicities as response variables, time-homogeneous transitions and $age$ at randomization as covariate affecting the initial probabilities of the latent variables.}
\end{figure}

Despite the relatively small sample size,  results showed stability among different models (see Table \ref{t:LMsel}).
The unrestricted LM model without covariates (M1) with the minimum BIC (16728.90) was obtained for $k=4$, identifying a latent process with four \textit{LOTox states}. Moreover, the basic model M2 with initial and transition probabilities parametrized by multinomial logit was preferable (BIC = 16512.16) to the unrestricted model M1 with the same number of latent states. Several models (M3-M12) with four latent states, obtained from M2 adding covariates effect to initial and/or transition probabilities, were fitted. By comparing models M3-M12 with M2, age (centred with respect to the mean) at randomization was the only covariate leading to a significant improvement in terms of both BIC and AIC (M5). Model M5, whose path diagram for a given subject $i$ is shown in Figure \ref{fig:M5}, was then selected as final model:
\begin{itemize}
	\item initial probabilities were associated with patient's \textit{age} at randomization and Equations (\ref{eq:betas}) for a patient $i$ became
	\begin{equation}\label{eq:M5delta}
		\log \frac{\delta_{u \mid age_i}}{\delta_{1 \mid age_i}} = \beta_{0 u}+ \beta_{1 u} \cdot (age_i - 15) \qquad u=2,3,4;
	\end{equation}
	\item transition probabilities were assumed time-homogeneous and Equations (\ref{eq:gammas}) became
	\begin{equation}\label{eq:M5tau}
		\log \frac{\tau_{u \mid \bar{u}}}{\tau_{\bar{u} \mid \bar{u} }} = \gamma_{0 \bar{u} u} \qquad \bar{u},u=1,2,3,4 \text{ with } \bar{u} \neq u.
	\end{equation}
\end{itemize}

\begin{table}[!h]
	\caption{\label{t:LMsel} Results for Latent Markov (LM) model selection for longitudinal toxicity data with different values of latent states $k$ and different
		restrictions. The maximum log-likelihood of each model is denoted by $\hat \ell$ and $g$ is the number of free parameters. WBC, PLT and NEUT in models M10-12 refers to white blood cell, platelets and neutrophils counts, respectively.}
	\small \centering
	\begin{tabular}{lrrrrr}
		\hline
		\multicolumn{1}{l}{\textbf{Latent Markov (LM) model}} & \multicolumn{1}{c}{$k$} & 
		\multicolumn{1}{c}{$g$} & \multicolumn{1}{c}{$\hat \ell$} & 
		\multicolumn{1}{c}{AIC} & 
		\multicolumn{1}{c}{BIC} \\ \hline
		M1: Unrestricted LM model without covariates &  1 & 18 & -8794.91 & 17625.81 & 17696.59 \\ 
		& 2 & 35 & -8420.19 & 16910.38 & 17048.01 \\ 
		& 3 & 68 & -8216.99 & 16569.98 & 16837.37 \\ 
		& 4 & 111 & -8035.21 & 16292.42 & 16728.90 \\ 
		& 5 & 164 & -7902.59 & 16133.18 & 16778.07 \\ 
		& 6 & 227 & -7793.14 & 16040.29 & 16932.91 \\ 
		& 7 & 300 & -7688.12 & 15976.24 & 17155.91 \\ 
		& 8 & 383 & -7603.30 & 15972.61 & 17478.66 \\ 
		& 9 & 476 & -7530.49 & 16012.98 & 17884.73 \\ 
		& 10 & 579 & -7462.34 & 16082.68 & 18359.45 \\ \hline
		M2: Multinomial logit LM model without covariates & 4 & 63 & -8069.21 & 16264.43 & 16512.16\\
		\hline
		M3: M2 + regimen effect on initial prob. & 4 & 66 & -8065.49 & 16262.97 & 16522.50 \\
		M4: M2 + gender effect on initial prob. & 4 & 66 & -8061.73 & 16255.45 & 16514.98\\
		M5: M2 + age effect on initial prob. & 4 & 66 & -8055.35 & 16242.69 & 16502.22\\ 
		\hline			
		M6: M2 + regimen effect on transition prob. & 4 & 75 & -8063.37 & 16276.74 &  16571.66 \\
		M7: M2 + gender effect on transition prob.& 4 & 75 & -8060.33 & 16270.66 & 16565.58\\
		M8: M2 + age effect on transition prob. & 4 & 75 & -8061.07 & 16272.14 & 16567.06\\	
		\hline
		M9: M2 + time-var chemotherapy dose on both prob. & 4 & 78 & -8045.55 & 16247.10 & 16553.82\\	
		M10: M2 + time-var WBC count on both prob. & 4 & 78 & -8062.53 & 16281.07 & 16587.78\\	
		M11:  M2 + time-var PLT count on both prob. & 4 & 78 & -8047.15 & 16250.30 & 16557.02\\	
		M12:  M2 + time-var NEUT count on both prob. & 4 & 78 & -8062.67 & 16281.34 & 16588.05\\	
		\hline
	\end{tabular}
\end{table}

\begin{figure}
	\centering
	\includegraphics[width=1\textwidth]{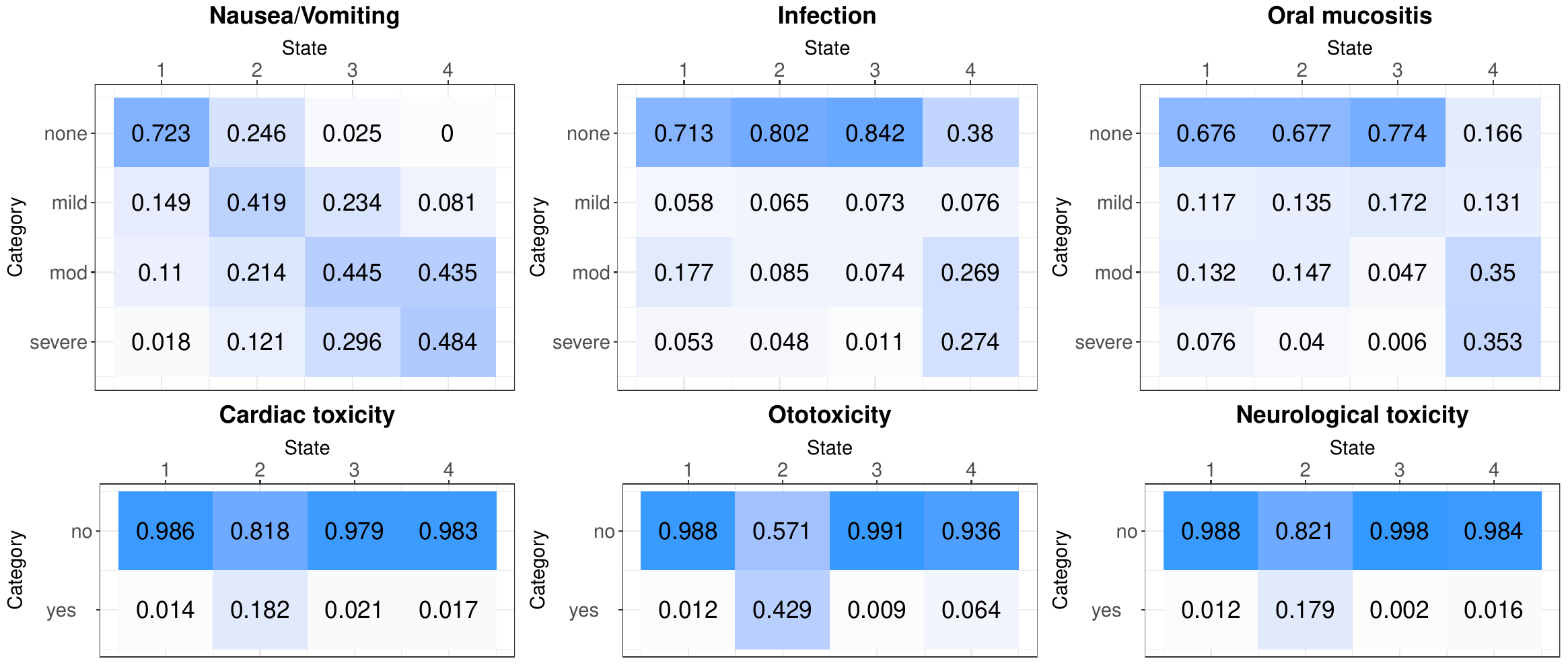}
	\caption{\label{fig:itemprob} Estimated conditional response probabilities $\hat{\phi}_{jy|u}$ for the final LM model in Figure \ref{fig:M5}. Each panel refers to a different toxicity $j \in \mathcal{J}=\{1: naus, \,2:inf, \,3:oral, \,4:car, \,5:oto, \,6:neur\}$. Each row refers to a response categories $y \in \{none; \,mild; \,moderate; \,severe\}$ for $j=1,2,3$ (\textit{generic} toxicities) and $y \in\{no; \,yes\}$ for $j=4,5,6$  (\textit{drug-specific} toxicities). Each column refers to a latent states $u\in\{1,2,3,4\}$.}
\end{figure}

Figure \ref{fig:itemprob} shows the estimated conditional response probabilities $\hat{\phi}_{jy|u}$ for each type of non-haematological toxicity under the selected model M5, which can be used for interpreting the latent states. In each toxicity-panel, each column refers to a different latent state $u\in\{1,2,3,4\}$. 
People in good conditions are allocated in state 1, since for all non-haematological toxicities the most probable category was the absence of the adverse event. 
State 2 seems to correspond to patients with non-severe nausea and it was the only state where \textit{drug-specific} toxicities occurred with a relevant probability, especially for ototoxicity where $\hat{\phi}_{51|2}=0.429$. 
State 3 seems to be characterized by patients undergoing only nausea or vomiting, mostly moderate or severe. 
In State 4 people with multiple \textit{generic} toxicities - mostly severe or moderate - with the certainty of having nausea ($\hat{\phi}_{10|4}=0$) are present. 
Based on these results, the following LOTox states labelling were derived:
\begin{itemize}
	\item State 1:\, quite good conditions (non-toxic)	$\rightarrow$ \textit{no LOTox}
	\item State 2:\, non-severe nausea with possible \textit{drug-specific} AEs $\rightarrow$ \textit{moderate LOTox}
	\item State 3:\, moderate/severe nausea/vomiting only $\rightarrow$ \textit{low LOTox} (limited to nausea)
	\item State 4:\, multiple severe/moderate \textit{generic} toxicities $\rightarrow$ \textit{high LOTox}.
\end{itemize}
Note that the states numbering (from 1 to 4) does not correspond with the progressive severity of overall toxicity burden (from \textit{no} to \textit{high}).

Table \ref{t:param} displays the estimated regression parameters  $\bm{\hat \beta}_u=\left(\hat \beta_{0u},\hat \beta_{1u}\right)$ for the initial probabilities in Equation (\ref{eq:M5delta}) and  the estimated transition probabilities $\hat \tau_{u \mid \bar{u}}$ in Equation (\ref{eq:M5tau}).
The estimated intercepts indicates that for 15-year patients the most prevalent state at cycle 1 was \textit{low LOTox} state 3 (limited to nausea), followed by \textit{no LOTox} state 1, \textit{high LOTox} state 4 and \textit{moderate LOTox} state 2. The estimates for $age$ were all positive, indicating that older individuals reported a higher overall severity at the first cycle compared to younger patients.
The estimated transition probabilities  $\hat \tau_{u \mid \bar{u}}$  shows a quite high persistence in the same state, especially for non-toxic state 1 and moderate state 2, where \textit{drug-specific AEs} may also lead to permanent conditions (see Appendix Table \ref{t:ctcae}). The highest transition probability was 15.6\% and was observed from the \textit{high LOTox} state 4, where the effects of \textit{generic AEs} are reversible and temporary, to the first non-toxic state.  Other transitions were observed from \textit{high LOTox} state 4 to nausea/vomiting only in state 3 (8.7\%) and from \textit{low LOTox} state 3 (limited to nausea) to \textit{no LOTox} state 1 (10.7\%) or \textit{high LOTox} state 4 (8.2\%).
The remaining transition probabilities were always lower than 8\%.

\begin{table}
	\caption{\label{t:param} Estimated regression parameters affecting the distribution of the initial probabilities in Equation (\ref{eq:M5delta}) and estimated transition probabilities in Equation (\ref{eq:M5tau}).}
	\centering \small
	\begin{tabular}{ccccc}
		\hline
		\multicolumn{5}{c}{Regression parameters for initial probabilities}\\ \hline
		& $u$ & 2 & 3 & 4 \\ 
		\hline
		Intercept & $\hat \beta_{0u}$ & -1.2679 & 1.0138 & -0.3031 \\ 
		Age & $\hat \beta_{1u}$ & 0.1858 & 0.0014 & 0.0512 \\ 
		\hline
		\hline
		\multicolumn{5}{c}{Transition probabilities from $\bar{u}$ to $u$ ($\hat \tau_{u \mid \bar{u}}$)}\\
		\hline
		$\bar{u} \setminus u $ & 1 & 2 & 3 & 4 \\ 
		\hline
		1 & 0.9674 & 0.0167 & 0.0032 & 0.0127 \\ 
		2 & 0.0525 & 0.9214 & 0.0245 & 0.0016 \\ 
		3 & 0.1070 & 0.0526 & 0.7581 & 0.0824 \\ 
		4 & 0.1555 & 0.0356 & 0.0868 & 0.7221 \\ 
		\hline
	\end{tabular}
\end{table}

\begin{figure}[t]
	\centering
	\includegraphics[width=0.85\textwidth]{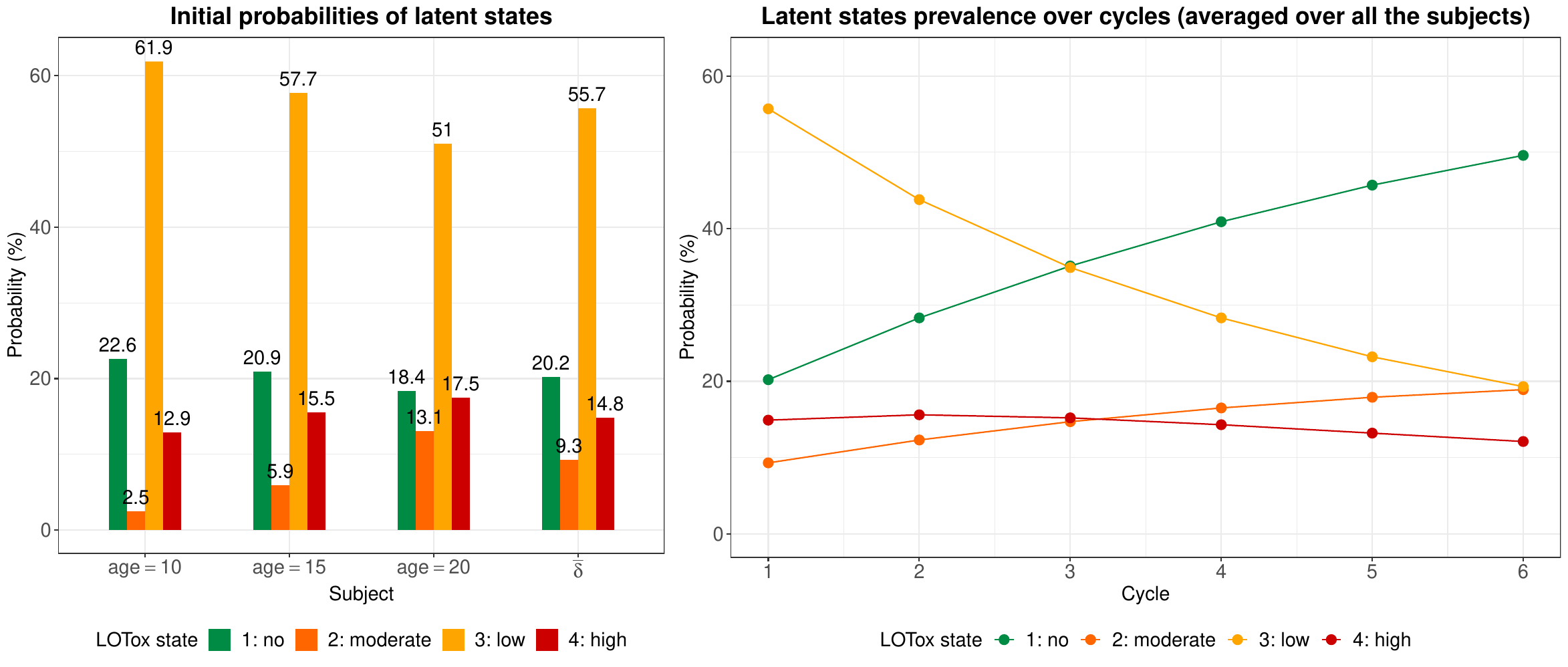}
	\caption{\label{fig:delta_tau} Left panel: estimated initial probabilities of latent states for patients aged 10, 15 and 20 years old and average $\bm{\bar\delta}$ of the initial probabilities over all the 377 subjects in the sample.\\ Right panel: latent states prevalences over cycles $t=1,\dots,6$ averaged over all the subjects.\\ Different colours refer to different Latent Overall Toxicity (LOTox) state (green: \textit{no LOTox state 1}; yellow: \textit{low LOTox state 3}; orange: \textit{moderate LOTox state 2};  red: \textit{high LOTox state 4}).}
\end{figure}

Starting from these parameter estimates, Figure \ref{fig:delta_tau} (left panel) displays the estimated vectors of initial probabilities $\bm{\hat\delta}_i = \left(\hat\delta_{1\mid age_i},\hat\delta_{2\mid age_i},\hat\delta_{3\mid age_i},\hat\delta_{4\mid age_i}\right)$ for patients aged 10, 15 and 20 years old and the vector $\bm{\bar\delta}= \left(\bar\delta_1,\bar\delta_2,\bar\delta_3,\bar\delta_4\right) = \left(0.202, 0.093, 0.557, 0.148\right)$ obtained as average of vectors $\bm{\hat\delta}_i$ over all the 377 subjects in the sample. On average, at cycle 1 \textit{low LOTox} state 3 of subjects with nausea/vomiting only had the largest dimension (55.7\%), followed by 20.2\% of individuals for \textit{no LOTox} state 1. \textit{No} and \textit{low LOTox} states together, representing the states with the lowest overall toxic severity, accounted for more than 75\% of the patients, whereas less than 25\% belonged to the latent states corresponding to the worst toxic conditions (\textit{moderate} and \textit{high LOTox} states 2 and 4).

Right panel in Figure \ref{fig:delta_tau} shows the estimated average probability of each latent state at each time-occasion, i.e., the latent states prevalences averaged over all the subjects at each cycle. On average, the presence of \textit{low} overall severity limited to nausea (state 3) decreased over cycles from 55.7\% to 19.3\% ($t=6$), whereas \textit{no} and \textit{moderate} overall toxicity (state 1 and 2, respectively) increased from 20.2\% to 49.7\% and from 9.2\% to 18.9\%. The presence in \textit{high} overall severity (state 4) was rather stable over cycles ranging in 10.1\%-15.6\%, with peaks at cycles 2 and 3.

\subsection{Longitudinal profiles of Latent Overall Toxicity}\label{s:results:profiles}
Once the parameters were estimated for the final LM model, the longitudinal latent probability profiles $\bm{p}_{iu}$ were reconstructed for each patient $i$ and latent state $u$, as explained in Section \ref{s:methods:prof}.
In case of longitudinal toxicity data, profiles $\bm{p}_{iu}$ in (\ref{eq:llpp}) are defined as \textit{longitudinal Probability profiles of LOTox} (\textit{P-LOTox}) since they represent the probability over cycles $t=1,2...,6$ of being in the LOTox state $u\in\{1,2,3,4\}$ for each patient $i$, given the observed toxic categories over treatment and individual characteristics (i.e., the age at randomization). 

\begin{figure}[!h]
	\centering
	\includegraphics[width=1\textwidth]{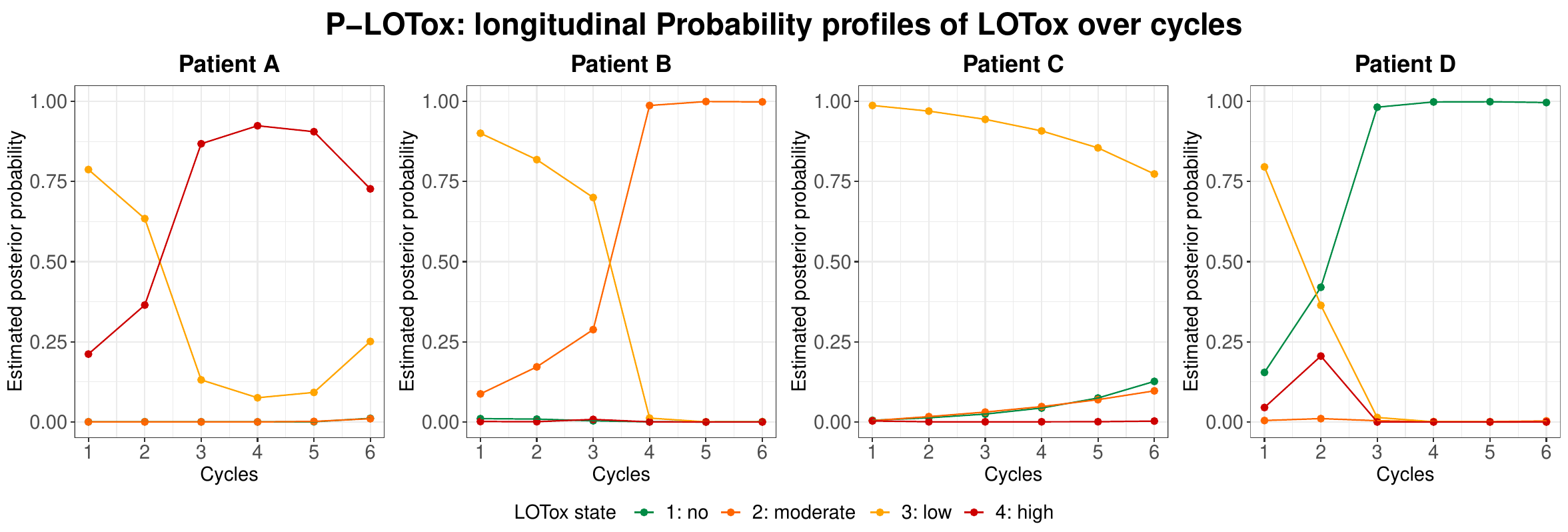}
	\caption{\label{fig:profiles} Longitudinal Probability profiles of Latent Overall Toxicity (P-LOTox) $\bm{p}_{iu}$. Each panel refers to a different patient $i=\{A,B,C,D\}$ in Appendix \ref{app:ptsAD}. Different colours refer to different latent states $u\in\{1,2,3,4\}$ (green: \textit{no LOTox state 1}; yellow: \textit{low LOTox state 3}; orange: \textit{moderate LOTox state 2};  red: \textit{high LOTox state 4}).}
\end{figure}

Figure \ref{fig:profiles} shows the longitudinal P-LOTox profiles $\bm{p}_{iu}$ for four patients $i=\{A,B,C,D\}$ aged 15 years old and with different observed toxic categories over cycles, as reported in Appendix \ref{app:ptsAD}. 
Each panel refers to a different patient and displays the individual realisations of the latent process over cycles. Different patterns of overall toxicity evolution during treatment can be observed between subjects, based on patient-specific observed toxicity data.
For example, right panel shows that at cycle 1 patient $D$ had probabilities 79.6\% of being in \textit{low LOTox} state, 15.5\% of having a non-toxic condition, 4.5\% and 0.4\% of \textit{high} and \textit{moderate LOTox}, respectively. The probabilities evolved over the cycles, as shown by the four profiles, ending with a 99.7\% probability of being in quite good conditions at the end of treatment. 

The lowest toxic burden is represented by the non-toxic state 1 of patients in good conditions, chosen as reference state ($R=1$: \textit{no LOTox}) to reconstruct the longitudinal latent relative risk profiles $\bm{r}_{iu}$ for each patient $i$ and latent state $u \in \{2,3,4\}$.
In case of longitudinal toxicity data, profiles $\bm{r}_{iu}$ in Equation (\ref{eq:lrr}) can be also called \textit{longitudinal Relative Risk profiles of LOTox} (\textit{RR-LOTox}). They represent for each patient $i$ the relative risk (in logarithmic scale) over cycles $t=1,2...,6$ of being in the LOTox state $u\in\{2,3,4\}$ rather than in the non-toxic state $R=1$, given the observed toxic categories over treatment and individual characteristics.

\begin{figure}[!h]
	\centering
	\includegraphics[width=1\textwidth]{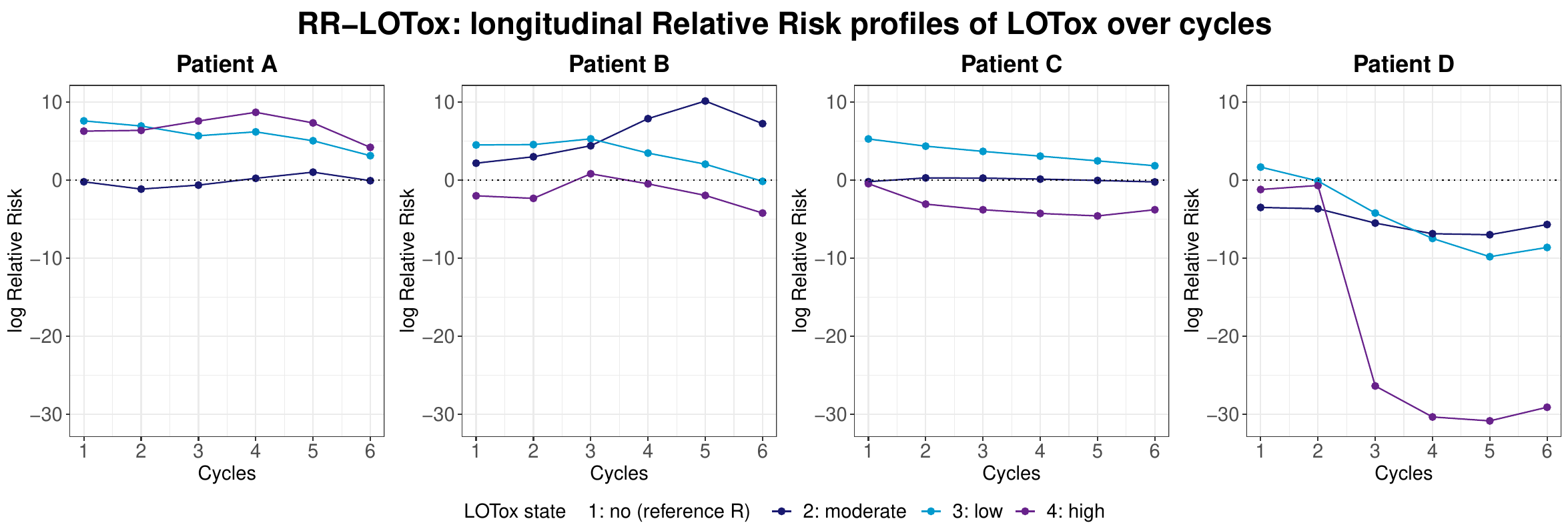}
	\caption{\label{fig:risks} Longitudinal Relative Risk profiles of Latent Overall Toxicity (RR-LOTox) $\bm{r}_{iu}$. Each panel refers to a different patient $i=\{A,B,C,D\}$ in Appendix \ref{app:ptsAD}. Reference LOTox state is \textit{no LOTox} state $R=1$. Different colours refer to different non-reference latent states $u\in\{2,3,4\}$ (light-blue: \textit{low LOTox state 3} vs \textit{no LOTox}; blue: \textit{moderate LOTox state 2} vs \textit{no LOTox};  purple: \textit{high LOTox state 4} vs \textit{no LOTox}).}
\end{figure}

Figure \ref{fig:risks} shows the longitudinal RR-LOTox profiles $\bm{r}_{iu}$ for patients $i=\{A,B,C,D\}$. Different toxic risk progressions during treatment can be observed among patients, depending on their observed toxicity data. For example, right panel shows that at first cycle patient $D$'s risk of being in \textit{low LOTox} state was 5.14 times higher the risk of having a non-toxic condition, whereas risks of \textit{high} and \textit{moderate LOTox} were 0.29 and 0.03 times lower, respectively. Then, RR-LOTox profiles evolved over the cycles, as shown by the four trajectories, ending up with negligible relative risks ($<0.01$) for \textit{low/moderate/high LOTox} conditions compared with a non-toxic condition at the end of treatment.

Both longitudinal P-LOTox and RR-LOTox profiles summarize and quantify the overall toxic risk over time for each patient based on observed individual characteristics, capturing differences in the overall history of toxicity across patients. P-LOTox profiles reflect the absolute size of the probabilities over time for each latent state, whereas RR-LOTox profiles focus on the relative risk with respect to the clinically desirable condition, i.e., the non-toxic one. 


\section{Discussion}
\label{s:discuss}
Due to the presence of multiple types of Adverse Events (AEs) with different levels of severity, identifying the actual extent of toxic burden and investigating the evolution of patient's overall toxicity represent challenging problems in cancer research. 
AEs are one of the main factors determining clinical decisions in medical interventions and treatment planning, playing a fundamental role in health assessment and patient monitoring. 
The development of statistical methods able to summarize multiple AEs and to deal with the complexity of chemotherapy data, considering both the longitudinal and categorical aspects of toxicity levels progression, is necessary and clinically relevant.

This paper proposed a new taxonomy based on LM models with covariates and CoDa methods to provide novel techniques for investigating the evolution of the latent overall toxicity condition for each patient over chemotherapy treatment. The novel approach was applied to longitudinal chemotherapy data from MRC BO06/EORTC 80931 Randomized Controlled Trial for osteosarcoma, considering patients who had successfully completed the six-cycle treatment. According to clinicians, the analysis of that specific patient cohort may cast some light on a complex aspect of this rare disease. It may indeed provide insights for the identification of novel evidence-based guidelines for an effective management of adverse symptoms and treatment. This is also relevant for the future development of innovative tools to support clinical decision-making in personalized interventions.

By assuming the existence of a LM chain for the LOTox condition of a patient, the proposed taxonomy identified sub-populations of patients characterized by a common distribution of toxic categories, and by a similar overall toxicity burden. Four LOTox states were found, which represent different levels of multiple AEs severity: (i) people in quite good conditions (\textit{no LOTox state 1}), (ii) patients undergoing only nausea or vomiting - mostly moderate or severe - (\textit{low LOTox state 3}), (iii) subjects with non-severe nausea and the possibility to develop \textit{drug-specific} AEs (\textit{moderate LOTox state 2}), or (iv) people with multiple severe/moderate \textit{generic} toxicities (\textit{high LOTox state 4}).
The LM approach estimated the initial prevalence of each state and the probability of individual changes over time. This allowed to reconstruct the patient-specific longitudinal LOTox profiles to assess the dynamic evolution of overall toxicity burden during treatment for each subject.

Both longitudinal P-LOTox and RR-LOTox profiles captured the individual realisations of the latent process over cycles, showing different patterns of overall toxicity evolution during treatment among patients. P-LOTox profiles illustrated the latent process using absolute terms, giving insights into the actual probabilities of being in the various LOTox states over cycles.  RR-LOTox profiles -- obtained by additive log-ratios transformation -- reported relative risk measures to emphasize the difference between low/moderate/high LOTox states and the clinically desirable non-toxic condition. These aspects can not be investigated using a simple path prediction (see Appendix \ref{app:decoding}). Together, absolute probabilities and relative risks provide a full picture of the individual LOTox dynamics during treatment, which may be considered as a proxy for patient’s quality of life and used to describe patient’s response to therapy over cycles in terms of toxic AEs.

The developed LM approach represents a general and flexible method to quantify the personal evolution of overall toxic risk during chemotherapy.
It can be adapted and applied to other cancer studies, even if characterized by different types of treatments (e.g., immunotherapies, molecularly targeted agents or radiation therapy) and thus by different adverse events. Indeed, the taxonomy can be redesigned for a different cancer study in the presence of  two main components: (i) a detailed protocol and/or collaboration with medical staff to identify treatment timing, relevant adverse events and patient's history; (ii) a proper longitudinal storage of toxicity data according to the CTCAE scale or an analogous pre-defined grading system.
A detailed description of how to adapt the proposed LM taxonomy to different tumour/treatment types can be found in Appendix \ref{app:cancer}.

This work opens doors to further research, both in the field of statistical methodology as well as in cancer research.
The additive log-ratios transformation allowed to remove non-negative and sum-to-one constraints of the CoDa vectors, mapping the compositions to a real space.
Standard statistical techniques could then be applied to the transformed data. 
Based on their different LOTox dynamics, patients could be stratified in different risk groups to be used during treatment, and association with survival outcomes may be investigated to provide new insights in the treatment effect during the evolution of the disease. 
Moreover, the proposed analysis could be further extended to include (i) informative drop-outs (e.g., discontinuation due to excessive toxicity, which depends on the latent variables); (ii) multi-state drop-outs (e.g., discontinuation for different reasons); (iii) intermittent missing responses between cycles. These extensions are not trivial and require both the development of complex methodologies and accurate data.

A novel approach is provided to summarise and quantify patient’s overall toxic risk and its evolution during treatment. 
This work also highlights the importance of a longitudinal, detailed and protocol-regulated collection of cancer treatment-toxicity data.
In cooperation with medical staff, this novel methodology 
might provide insights for the definition of new guidelines to reduce the impact of cancer treatments in terms of toxicity burden.

\small
\subsection*{\small Acknowledgements}
The authors gratefully thank MRC Clinical Trials Unit at UCL, Institute of Clinical Trials and Methodology, UCL, London for making data from MRC BO06/EORTC 80931 trial  available for this research. 

\subsection*{\small Data Availability Statement}
Data are not publicly available due to privacy restrictions.  Access to the full dataset of MRC BO06 trial can be requested to MRC Clinical Trials Unit at UCL, Institute of Clinical Trials and Methodology, UCL, London.

\small
%\bibliographystyle{apalike}
%\bibliography{bibliography}

\normalsize

\newpage

\appendix
\section*{Appendix}

\renewcommand\thesubsection{A}
\subsection{\large MRC BO06/EORTC 80931 RCT protocol}
\label{app:protocol}

\begin{figure}[h!]
	\centering
	\includegraphics[width=1\textwidth]{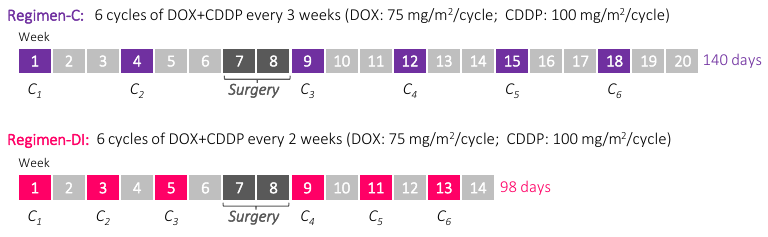}
	\renewcommand\thefigure{A.1}
	\caption{Patients are randomized at baseline to one of the two regimens, with the same anticipated cumulative dose but different duration.}
	\label{fig:reg}
\end{figure}

Data from the MRC BO06/EORTC 80931 Randomized Controlled Trial (RCT) for patients with non-metastatic high-grade osteosarcoma recruited between 1993 and 2002 were analysed (\citealp{lewis2007}). The trial randomised patients between conventional treatment with doxorubicin (DOX) and cisplatin (CDDP) given every 3 weeks (\textit{Reg-C}) versus a dose-intense regimen of the same two drugs given every 2 weeks (\textit{Reg-DI}), supported by granulocyte colony-stimulating factor. Chemotherapy was administered for six cycles (a cycle is a period of either 2 or 3 weeks depending on the allocated regimen), before and after surgical removal of the primary osteosarcoma. In both arms, DOX (75  mg/m\textsuperscript{2}) plus CDDP (100  mg/m\textsuperscript{2}) were given over six cycles. Surgery to remove the primary tumour was scheduled at week 6 after starting treatment in both arms, that is, after 2 cycles (2 $\times$ [DOX+CDDP]) in regimen-C and after 3 cycles (3 $\times$ [DOX+CDDP]) in regimen-DI. Postoperative chemotherapy was intended to resume 2 weeks after surgery in both arms. Figure \ref{fig:reg} shows the trial design. Laboratory tests were usually performed before each cycle of chemotherapy (in some cases also during and after the cycle) in order to monitor patient's health status. %and the development of toxicities or adverse events. 
For each completed cycle, non-haematological chemotherapy-induced toxicity for nausea/vomiting, infection, oral mucositis, cardiac toxicity, ototoxicity and neurological toxicity were graded according to the Common Terminology Criteria for Adverse Events Version 3 (CTCAE v3.0) by \cite{ctcae3}, with grades ranging from 0 (none) to 4 (life-threatening), as shown in Table \ref{t:ctcae}. Delays or chemotherapy dose reductions during treatment were possible in case of toxicity. 
In case of disease progression, excessive toxicity, refusal of chemotherapy or other complications, patient's treatment was terminated without completion of all six cycles.
Additional details can be found in the primary analysis of the trial by \citet{lewis2007}.

%\newpage
\begin{table}%[h!]
	\small
	\renewcommand\thetable{A.1}
	\caption{Toxicity coding based on Common Terminology Criteria for Adverse Events (CTCAE) v3.0 by \cite{ctcae3} for non-haematological chemotherapy-induced toxicity related to nausea/vomiting, infection, oral mucositis, cardiac toxicity, ototoxicity and neurological toxicity.}
	\label{t:ctcae}
	\begin{center}
		\begin{tabular}{p{0.16\textwidth}|p{0.1\textwidth}p{0.15\textwidth}p{0.15\textwidth}p{0.15\textwidth}p{0.15\textwidth}}
			\hline
			\textbf{Toxicity} & \textbf{Grade 0} & \textbf{Grade 1} &\textbf{Grade 2} &\textbf{Grade 3} &\textbf{Grade 4} \\ \hline
			Nausea/Vomiting & None & Nausea & Transient vomiting & Continuative vomiting  & Intractable vomiting\\ \hline
			Infection & None & Minor infection & Moderate infection  & Major infection & Major infection with hypotension\\ \hline
			Oral Mucositis & No change & Soreness or erythema & Ulcers: can eat solid & Ulcers: liquid diet only & Alimentation not possible\\ \hline
			Cardiac toxicity & No change & Sinus tachycardia & Unifocal PVC arrhythmia & Multifocal PVC & Ventricular tachycardia\\ \hline
			Ototoxicity & No change & Slight hearing loss & Moderate hearing loss & Major hearing loss & Complete hearing loss\\ \hline
			Neurological toxicity & None & Paraesthesia & Severe paraesthesia & Intolerable paraesthesia & Paralysis\\
			\hline
		\end{tabular}
	\end{center}
\end{table}

\renewcommand\thesubsection{B}
\subsection{\large Path prediction for latent Markov models}
\label{app:decoding}

In latent Markov models literature, once the model has been estimated, a decoding procedure is usually implemented to obtain a path prediction $\overset{*}{\bm u}_{i} = \left(\overset{*}{u}_{i}^{(1)},\dots,\overset{*}{u}_{i}^{(T)}\right)$ of the most likely latent states over time for each subject $i$, on the basis patient-specific observed data.

Among the developed procedures, \textit{local decoding} finds the most likely state occupied by a subject at any time point $t$: elements of $\overset{*}{\bm u}_{i}$ can be obtained by maximizing the \textit{posterior probabilities} at each time $t$ in Equation \eqref{eq:posterior}, as follows
\begin{equation*}
	\overset{*}{u}_{i}^{(t)} = \max_{u \in \{1,...,k\}} p_{iu}^{(t)} = \max_{u \in \{1,...,k\}} \mathrm{P}\left(U_i^{(t)}=u \big| \bm{\widetilde Y}_i=\bm{\widetilde y}_i, \bm{\widetilde X}_i=\bm{\widetilde x}_i\right) \quad \text{for all } t=1,...,T.
\end{equation*}

As an alternative, \textit{global decoding} finds the most likely sequence of latent states for a given subject on the basis of the responses he/she provided. It is based on an  adaptation of the Viterbi algorithm (\citealp{viterbi1967,juanrabiner1991}) which maximises the \textit{joint conditional probability} for each subject $i$, i.e.,
\begin{equation*}
	\overset{*}{\bm u}_{i} = \argmax_{\bm{u}} \, \mathrm{P}\left(\bm{U}_i=\bm{u} \big| \bm{\widetilde Y}_i=\bm{\widetilde y}_i, \bm{\widetilde X}_i=\bm{\widetilde x}_i\right),
\end{equation*}
through a forward-backward recursion. For further details see \citet{bartolucci2013book,bartolucci2014}.

\subsubsection{\normalsize \textit{Data Application: LOTox sequences}}

In case of longitudinal toxicity data, path prediction $\overset{*}{\bm u}_{i}$ represents the sequence of LOTox states over time for subject $i$. Let us consider the four patients aged 15 years old with different observed toxic categories over cycles reported in Appendix \ref{app:ptsAD}. The \textit{LOTox sequences} for patients $i=\{A,B,C,D\}$ can be then obtained as 

\begin{enumerate}
	\item[(i)] the sequences of the most probable LOTox states at each cycle $t$ (i.e., \textit{local decoding})
	\begin{equation*}
		\overset{*}{\bm u}_{A} = (3,3,4,4,4,4), \quad \overset{*}{\bm u}_{B} = (3,3,3,2,2,2), \quad
		\overset{*}{\bm u}_{C} = (3,3,3,3,3,3), \quad \overset{*}{\bm u}_{D} = (3,1,1,1,1,1);
	\end{equation*}
	\item[(ii)]  or the sequences of the most likely LOTox states across cycles (i.e., \textit{global decoding})
	\begin{equation*}
		\overset{*}{\bm u}_{A} = (3,3,4,4,4,4), \quad \overset{*}{\bm u}_{B} = (3,3,3,2,2,2), \quad
		\overset{*}{\bm u}_{C} = (3,3,3,3,3,3), \quad \overset{*}{\bm u}_{D} = (3,3,1,1,1,1).
	\end{equation*}
\end{enumerate}
Differences between (i) and (ii) are due to the different types of probabilities that are maximized, respectively posterior and joint conditional probabilities.
The individual \textit{LOTox sequence} allows to predict the LOTox state to which every patient belongs at a given cycle. However, it represent a summary of how the entire latent process evolves during treatment for a patient, as it only provides information about the most-likely condition without giving details about other states.

\renewcommand\thesubsection{C}
\subsection{\large Observed toxic categories over cycles for patients A--D}
\label{app:ptsAD}
Table \ref{t:profpts} reports the observed toxic categories over cycles related to four 15-year patients named A, B, C and D, whose relative \textit{longitudinal probability/relative risk profiles of LOTox} are shown in Figures \ref{fig:profiles} and \ref{fig:risks}, respectively .

\begin{table}[h!]
	\small \renewcommand\thetable{C.1}
	\caption{Observed toxicity categories over cycles $t=1,...,6$ for four random patients $i \in\{A,B,C,D\}$ aged 15 years old. Categories for \textit{generic} toxicities (nausea, infection and oral mucositis) are $\{0:none, \,1:mild, \,2:moderate, \,3:severe\}$ $(j=1,2,3)$. Categories for \textit{drug-specific} toxicities (cardiac toxicity, ototoxicity and neurological toxicity) are $\{0:no, \,1:yes\}$  $(j=4,5,6)$. For each patient $i$ the complete response vector is $\bm{\tilde{y}}_i = \left(\bm{{y}}_i^{(1)},\dots,\bm{{y}}_i^{(1)}\right)$ where $\bm{{y}}_i^{(t)}=\left(y_{i1}^{(t)},\dots,y_{i6}^{(t)}\right)$.}
	\label{t:profpts}
	\centering
	\begin{tabular}{ccccccccc}
		\hline
		Patient $i$& Cycle $t$ &  $age_i$ & Naus $y_{i1}^{(t)}$  & Inf $y_{i2}^{(t)}$ & Oral $y_{i3}^{(t)}$& Car $y_{i4}^{(t)}$ & Oto $y_{i5}^{(t)}$ & Neur $y_{i6}^{(t)}$ \\ 
		%$i$ & $t$ & $age_i$ &  & $y_{i2}^{(t)}$ & $y_{i3}^{(t)}$ &  $y_{i4}^{(t)}$ & $y_{i5}^{(t)}$ & $y_{i6}^{(t)}$ \\
		\hline
		A & 1 & 15 & 3 & 0 & 1 & 0 & 0 & 0 \\ 
		& 2 &  & 3 & 1 & 0 & 0 & 0 & 0 \\ 
		& 3 &  & 3 & 3 & 0 & 0 & 0 & 0 \\ 
		& 4 &  & 3 & 2 & 1 & 0 & 0 & 0 \\ 
		& 5 &  & 3 & 0 & 2 & 0 & 0 & 0 \\ 
		& 6 &  & 3 & 0 & 1 & 0 & 0 & 0 \\ 
		\hline
		B & 1 & 15 & 1 & 0 & 0 & 0 & 0 & 0 \\ 
		& 2 &  & 1 & 0 & 0 & 0 & 0 & 0 \\ 
		& 3 &  & 3 & 0 & 0 & 0 & 0 & 0 \\ 
		& 4 &  & 1 & 0 & 0 & 0 & 1 & 0 \\ 
		& 5 &  & 1 & 0 & 0 & 0 & 1 & 0 \\ 
		& 6 &  & 1 & 0 & 0 & 0 & 1 & 0 \\ 
		\hline
		C & 1 & 15 & 2 & 0 & 0 & 0 & 0 & 0 \\ 
		& 2 &  & 1 & 0 & 0 & 0 & 0 & 0 \\ 
		& 3 &  & 1 & 0 & 0 & 0 & 0 & 0 \\ 
		& 4 &  & 1 & 0 & 0 & 0 & 0 & 0 \\ 
		& 5 &  & 1 & 0 & 0 & 0 & 0 & 0 \\ 
		& 6 &  & 1 & 0 & 0 & 0 & 0 & 0 \\ 
		\hline
		D & 1 & 15 & 2 & 0 & 0 & 0 & 0 & 0 \\ 
		& 2 &  & 2 & 0 & 2 & 0 & 0 & 0 \\ 
		& 3 &  & 0 & 0 & 0 & 0 & 0 & 0 \\ 
		& 4 &  & 0 & 0 & 0 & 0 & 0 & 0 \\ 
		& 5 &  & 0 & 0 & 0 & 0 & 0 & 0 \\ 
		& 6 &  & 0 & 0 & 0 & 0 & 0 & 0 \\ 
		\hline
	\end{tabular}
\end{table}

\renewcommand\thesubsection{D}
\subsection{\large How to adapt the taxonomy to different cancer/treatment types}
\label{app:cancer}

The taxonomy developed in this work represents a general and flexible method to quantify the personal evolution of overall toxic risk during chemotherapy. Thanks to its flexibility, it can be adapted and applied to other cancer studies, even if characterized by different types of treatments (such as immunotherapies, molecularly targeted agents or radiation therapy) and thus by different adverse events.

The taxonomy can be redesigned for a different cancer/treatment study in presence of  two main components:
\begin{itemize}
	\item[(i)] a detailed protocol and/or collaboration with medical staff to identify 
	\begin{itemize}
		\item treatment timing,
		\item relevant adverse events,
		\item patient's history;
	\end{itemize}
	\item[(ii)] longitudinal information about the toxicity data, recorded according to the CTCAE scale (\citealp{ctcae3}) or a similar pre-defined grading system.
\end{itemize}
With regard to the notation of the Latent Markov (LM) model presented in Section \ref{s:methods:model}, (i) is fundamental for identifying
\begin{itemize}
	\item the time points $t=1,\dots,T$, which could be cycles, visits, therapy sessions, etc,
	\item the set of categorical response variable $\mathcal{J}$ and relative toxicities $j=1,\dots,J$,
	\item the complete vector of individual covariates $\widetilde{\boldsymbol{X}}_i$ for each patient $i$ (i.e., patient's history),
\end{itemize}
whereas (ii) is necessary for defining
\begin{itemize}
	\item the set $\mathcal{C}_j$ of toxic categories for each toxicity $j$, obtained possibly merging low-frequency classes (see Section \ref{s:data:long});
	\item the complete response vector $\widetilde{\boldsymbol{Y}}_i$ of observed toxic categories for each patient $i$.
\end{itemize}
The above characteristics must be selected on the basis of relevance to the cancer/treatment under study.  Moreover,  the response toxicity variables must be conditionally independent, as in the taxonomy each observed response is expected to depend only on the corresponding “true'' Latent Overall Toxicity (LOTox) level (see Sections \ref{s:methods:lmmot} and \ref{s:methods:model}). 

Once these aspects have been identified and defined, the LOTox condition of each patient can be assumed as the latent status of interest and model selection can be performed to identify the final LM model (see Section \ref{s:methods:sel}). Finally, longitudinal Probability profiles of Latent Overall Toxicity (P-LOTox) and longitudinal Relative Risk profiles of Latent Overall Toxicity (RR-LOTox) can be reconstructed for each patient (see Section \ref{s:methods:prof}). 

This process of model refitting to other studies requires longitudinal, detailed and protocol-regulated cancer treatment data, highlighting the importance of an accurate data collection.

\end{document}